\documentclass[useAMS,usenatbib]{mn2e}
\usepackage{aas_macros}
\usepackage[dvips]{graphicx}
\usepackage{amssymb}
\title[Environmental Dependence of Halo Properties at High Redshifts]{The Evolution of Dark Matter Halo Properties  in Clusters, Filaments, Sheets and Voids}
\author[O. Hahn et al.]{Oliver Hahn,$^{1}$\thanks{E-mail: hahn@phys.ethz.ch}
C. Marcella Carollo,$^{1}$ Cristiano Porciani$^{1}$ and Avishai Dekel$^{2}$\\
$^{1}$ETH Z\"urich, 8093 Z\"urich, Switzerland\\
$^{2}$Racah Institute of Physics, The Hebrew University, Jerusalem 91904, Israel }

\begin{document}

\date{MNRAS in press}
\pagerange{\pageref{firstpage}--\pageref{lastpage}} \pubyear{2007}
\maketitle

\label{firstpage}

\begin{abstract}
We use a series of high-resolution N-body simulations of the concordance cosmology to investigate the redshift evolution since $z=1$ of the properties and alignment with the large-scale structure of haloes in clusters, filaments, sheets and voids. We find that: $(i)$ Once a rescaling of the halo  mass  with $M_\ast(z)$, the typical mass scale collapsing at redshift $z$, is performed, there is no further significant redshift dependence in the halo properties; $(ii)$ The environment influences the halo shape and formation time at all investigated redshifts for haloes with masses $M\lesssim M_\ast$; and $(iii)$ There is a significant alignment of both spin and shape of haloes with filaments and sheets. In detail, at all redshifts up to $z=1$:  {\it a)} Haloes with masses below  $\sim M_\ast$ tend to be more oblate when located in clusters than in the other environments; this trend is reversed at higher masses: above about $M_\ast$, halos in clusters are typically more prolate than similar massive halos in sheets, filaments and voids. {\it b)} The haloes with $M\gtrsim M_\ast$ in filaments spin more rapidly than similar mass haloes in clusters; haloes in voids have the lowest median spin parameters; {\it c)} Haloes with $M\lesssim M_\ast$ tend to be younger in voids and older in clusters; {\it d)} In sheets, halo spin vectors tend to lie preferentially within the sheet plane independent of halo mass; in filaments, instead, haloes with $M\lesssim M_\ast$ tend to spin parallel to the filament and higher mass haloes perpendicular to it. For halo masses $M\gtrsim M_\ast$, the major axis of haloes in filaments and sheets is strongly aligned with the host filament or the sheet plane, respectively. Such halo-LSS alignments may be of importance in weak lensing analyses of cosmic shear. A question that is opened by our study is why, in the $0 < z < 1$ redshift regime that we have investigated, the mass scale for gravitational collapse, $M_\ast$, sets  roughly the threshold below which the large-scale structure environment either begins to affect, or reverses,  fundamental properties of dark matter haloes.
 \end{abstract}


\section{Introduction}
Numerical simulations of concordance cosmology have shown that properties of dark matter haloes do not depend only on the mass of the halo, as suggested by prior analytical work based on the excursion set theory for structure evolution \citep[e.g.][]{Bond1991,Lacey1993}. Rather, they also depend on the environment in which the halo resides \citep[e.g.][Paper I in the following]{ Gao2005,Wechsler05,Harker2006,Wang06,Hahn06}. This dependence on environment is quite significant at redshift zero for low mass haloes, typically with masses $<5\times10^{12}\,h^{-1}{\rm M}_\odot$. 
In detail, using marked statistics, \cite{Sheth2004} find evidence that haloes of a given mass form earlier in dense regions. High resolution simulations confirm this finding and quantify it as a function of halo mass \citep{Gao2005,Croton06,Harker2006,Reed06,Maulbetsch06}.
At the same time it has become clear that also other halo properties as concentration and spin correlate
with local environment \citep{AvilaReese2005,Wechsler05,Bett06,Maccio06,Wetzel06,Hahn06}.
\cite{Gao06} find that haloes with, e.g., high spin-parameter or formation time, tend to be more strongly clustered than younger and low-spin haloes. It is possible that this environmental dependence of halo properties has also an impact on the baryonic galaxies. Galaxy properties in the local Universe are known to vary systematically with environment \citep[e.g.][]{Dressler80, Kauffmann04, Blanton05}. 

In this paper, we extend the investigation of the properties of dark matter haloes as a function of environment to high redshifts. In particular, we follow the definition of environment that we presented in Paper I, which associates haloes to four classes with different dynamical properties: voids, sheets, filaments and clusters. These four environments are identified on the basis of a tidal stability criterion for test particles which is inspired by the Zel'dovich approximation \citep{Zeldovich70}. We find that, at each redshift, all investigated properties of haloes show some correlation with mass and environment, and that the redshift dependence of halo properties with mass is removed when such properties are investigated as a function of the rescaled mass $M/M_\ast$, where $M_\ast$ is the typical mass-scale collapsing at each epoch. 

The paper is organised as follows. In Section \ref{sec:Sims}, we briefly summarise the specifics of our N-body simulations, the definitions of the halo properties that we study and the definitions of the four environments - clusters, sheets, filaments and voids. We present the results in Section \ref{sec:Haloes} and summarise our conclusions in Section \ref{sec:Conclusion}.

\section{Numerical Simulations and Definitions}
\label{sec:Sims}
We use the three high-resolution cosmological N-body simulations described in more detail in Paper I, which were obtained  with the tree-PM code {\sc Gadget-2} \citep{Springel2005}. These simulations are used to follow the formation and evolution of large-scale structure in a flat $\Lambda$CDM cosmology with matter density parameter $\Omega_{\rm m}=0.25$, baryonic contribution $\Omega_{\rm b}=0.045$ and a present-day value of the Hubble constant $H_0=100\,h\,{\rm km}\,{\rm s}^{-1}\,{\rm Mpc}^{-1}$ with $h=0.73$ with an initial power spectrum normalised to have $\sigma_8=0.9$. Each simulation consists of $512^3$ collisionless dark matter particles in periodic boxes of sizes $L_1=45\,h^{-1}\,{\rm Mpc}$, $L_2=90\,h^{-1}\,{\rm Mpc}$ and $L_3=180\,h^{-1}\,{\rm Mpc}$, respectively. The corresponding particle masses are $4.7\times 10^7$, $3.8\times 10^8$ and $3.0\times 10^9\,h^{-1}{\rm M}_\odot$ for the three boxes. Initial conditions were generated using the {\sc Grafic2} tool \citep{Bertschinger2001}. Particle positions and velocities were saved for 30 time-steps, logarithmically spaced in expansion parameter $a$ between $z=10$ and $z=0$. The mass-range of these three simulations allows us to resolve haloes with masses below $M_\ast$ up to redshifts $z \lesssim 1$.

The halo properties that we investigate are formation redshift, shape, and spin parameter. The formation redshift and shape parameters are defined as in Paper I; we adopt however a slightly different approach to measure the halo spin parameter than in our previous work. We summarise our definitions below.

\subsection{Halo Catalogues}
\label{sec:HaloCatalogues}
Haloes were identified in each snapshot using the standard friends-of-friends \citep[FOF, ][]{Davis1985} algorithm with a linking length equal to $0.2$ times the mean inter-particle distance. Haloes that are well-resolved in each of the three simulations are then combined into one single catalogue. Unrelaxed systems were identified and deleted from the halo catalogues. These unrelaxed systems  are mainly contributed by close-pair halo configurations which are spuriously linked into one single halo. To identify them, we follow \cite{Bett06} and define the virialization parameter
\begin{equation}
\alpha \equiv \frac{2K}{V}+1,
\end{equation}
where $K$ is the total kinetic energy including the Hubble flow with respect to the centre of mass and $V$ is the total potential energy of the isolated FOF halo. The potential is computed using a tree for groups with more than $5000$ particles, and via direct summation for smaller halos. The virial theorem states that the time average of $\alpha$ vanishes for any isolated relaxed object. However, infalling material exerts a surface pressure such that $\alpha\lesssim 0$ \citep{Hetznecker06}. In addition, structures that are gravitationally bound have $\alpha > -3$. In order to exclude accidentally linked unvirialised structures or haloes that are just about to merge, it suffices to fix bounds on $\alpha$. In order to directly compare with \cite{Bett06}, we make the same choice of $|\alpha|<1/2$ that was adopted by those authors to set the threshold between virialized and non-virialized structures. 

Finally, we also exclude from our halo catalogues all those structures where the distance between the centre of mass $\mathbf{r}_{\rm CM}$ and the most bound particle $\mathbf{r}_{\rm MB}$ exceeds a fixed fraction $f=0.25$ of the largest distance between a particle in the halo and the centre of mass $\mathbf{r}_{\rm max}$, i.e. $f=|\mathbf{r}_{\rm CM}-\mathbf{r}_{\rm MB}|/|\mathbf{r}_{\rm max}|$. 

 The cleaning of the halo catalogues has a strong effect on the spin parameter distribution but only a minor influence on the other quantities that we study in this paper.

\subsection{Formation Redshift}

For each halo at redshift $z$, we identify a progenitor at $z_p>z$ by identifying particles that are contained in both haloes. The main progenitor is then chosen to be the most massive halo at each redshift that contributes at least 50 per cent of its particles to the final halo. We then define the formation redshift $z_{\rm form}$ to be the epoch at which a main progenitor which has at least half of the final mass first appears in the simulation; specifically,  $z_{\rm form}$ is found by linearly interpolating  between simulation snapshots in $\log z$ to find the point where half of the given halo mass is accumulated.

\subsection{Halo Shape}

In order to determine the shape of haloes, we use the moment of inertia tensor
\begin{equation}
I_{jk} \equiv m\sum_{i}\left(r_i^2\delta_{jk}-x_{i,j}x_{i,k}\right),
\end{equation}
where $m$ is the particle mass, $r_i\equiv |(x_{i,1},x_{i,2},x_{i,3})|$ is the distance of the $i$-th particle from the centre of mass of the halo and $\delta_{jk}$ denotes the Kronecker symbol. Given the lengths of the principal axes of inertia $l_1\geq l_2\geq l_3$, we then use the following definitions of sphericity $S$ and triaxiality $T$:
\begin{equation}
S = \frac{l_3}{l_1}\qquad \textrm{and}\qquad
T  = \frac{l_1^2-l_2^2}{l_1^2-l_3^2}.
\end{equation}
We find that a minimum of 500 particles per halo guarantees numerically reliable estimates of the shape parameters.

\subsection{Halo Spin Parameter}
We estimate the spin parameter \citep{Peebles1969} of a halo using the simplified form \citep{Bullock2001}
\begin{equation}
\lambda^{\prime} \equiv \frac{\left|\mathbf{J}_{\rm vir}\right|}{\sqrt{2}M_{\rm vir}V_{\rm vir} R_{\rm vir}}.
\end{equation}
Here all quantities with the subscript ``vir'' (angular momentum, mass and circular velocity) are computed within a sphere of radius $R_{\rm vir}$ around the most bound particle enclosing a mean density of $\Delta(z)\rho_{\rm c}(z)$, where $\rho_{\rm c}(z)$ is the critical density, and $\Delta(z)$ the density parameter according to the spherical collapse model. This density parameter can be approximated by \citep{Bryan1998}:
\begin{equation}
\Delta(z) = 18\,\pi^2+82\,f(z)-39\,f^2(z),
\end{equation}
with
\begin{equation}
f(z) = \frac{-\Omega_{\Lambda}}{\Omega_{\rm m}\,(1+z)^3+\Omega_{\Lambda}}
\end{equation}
in a flat cosmology. Under the assumption that the halo is in dynamical equilibrium, $V_{\rm vir}^{2}=GM_{\rm vir}/R_{\rm vir}$, 
the spin parameter can be rewritten as
\begin{equation}
\lambda^{\prime} = \frac{\left|\mathbf{J}_{\rm vir}\right|} {\sqrt{2GR_{\rm vir}} M_{\rm vir}^{3/2}}.
\end{equation}
Systematic numerical artefacts were found to be negligible for haloes consisting of at least 300 particles. The cleaning of the halo catalogues, as described in Section \ref{sec:HaloCatalogues}, has a strong influence on the distribution of $\lambda^{\prime}$. The probability for two haloes of similar mass to be erroneously linked by the halo finder grows both with the environmental density and decreasing mass of the haloes. This leads to an increasing component of unrelaxed structures of low mass in the spin distribution for which the virialisation conditions are not fulfilled. Furthermore, the angular momentum $\mathbf{J}$ is dominated by the orbital angular momentum of the pair rather than the intrinsic spin of either one of them. Exclusion of unrelaxed objects removes the tail of these apparent high-spin haloes with $\lambda^{\prime}\gtrsim 0.1$.

\subsection{Environmental Classification}
\label{sec:Classification}
\begin{figure*}
   \begin{center}
   \includegraphics[width=0.6\textwidth]{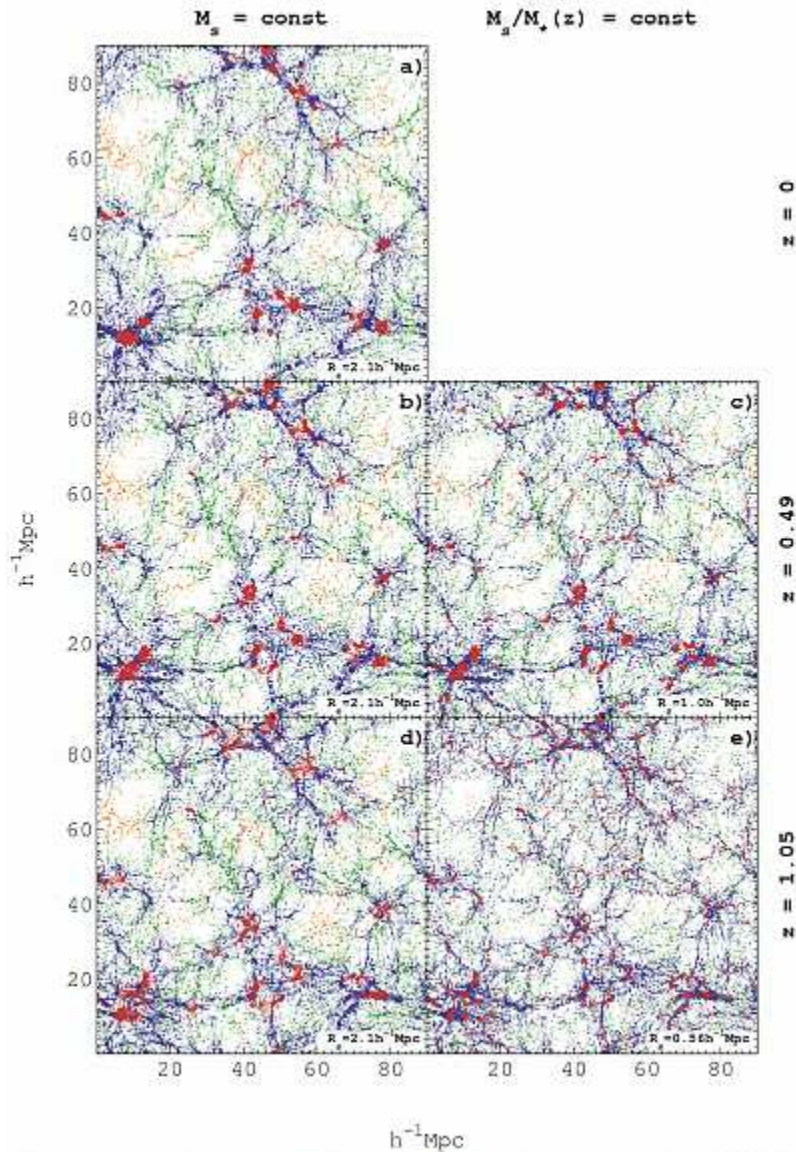}
 \end{center}
 \caption{ \label{fig:ScatterHaloes} Halo environment classification as a function of redshift in  projected slices of 15 $h^{-1}\,{\rm Mpc}$ for the 90 $h^{-1}{\rm Mpc}$ box. Only for presentation purposes, i.e., to increase the contrast in the figure, we represent with a dot  haloes  down to structures with only 10 particles.   The four environments are identified by colour: voids (orange), sheets (green), filaments (blue) and clusters (red). 
 Panel a) is the slice at $z=0$; the smoothing mass scale is $M_s=10^{13}\,h^{-1}\,{\rm M}_\odot$. Panels b) and c) are snapshots at  $z=0.49$; panels  c) and d) are at  $z=1.05$. In panels b) and d) the smoothing scale $M_s$ adopted for the environmental classification is as at $z=0$;  in panels c) and e) it is the ratio $M_s/M_{\ast}$ that is kept fixed.}
\end{figure*}
\label{sec:EnvironmentDefinition}

We employ the definition of environment that was described in detail in Paper I. This classification is based on a local stability criterion for the orbits of test particles in the matter distribution at a fixed epoch. A series expansion of the equation of motion for a test particle in the smoothed matter distribution gives a zero order acceleration and a first order deformation term that is determined by the tidal field tensor, i.e. the Hessian of the peculiar gravitational potential. We then classify the contraction/expansion behaviour of this first order term by the number of its positive/negative eigenvalues. In analogy with Zel'dovich theory \citep{Zeldovich70}, the application of this criterion provides four environmental categories defined by the signs of the three eigenvalues $\lambda_1\leq\lambda_2\leq\lambda_3$, namely: 
\begin{enumerate}
\item {\it clusters} ($\lambda_{1,2,3}\geq 0$), 
\item {\it filaments} ($\lambda_{2,3}\geq 0, \lambda_1<0$), 
\item {\it sheets} ($\lambda_3\geq 0, \lambda_{1,2}<0$), and 
\item {\it voids} ($\lambda_{1,2,3}<0$). 
\end{enumerate}
This definition relies on one free parameter, the length scale $R_s$ of the Gaussian filter that is used to smooth the matter distribution before obtaining the gravitational potential on a grid. As done in Paper I, we fix the smoothing scale at redshift $z=0$ to be $R_s=2.1\,h^{-1}\,{\rm Mpc}$, as this value provides the best agreement between the outcome of the orbit-stability criterion and our a-posteriori visual classification of the different environments. The smoothing length scale $R_s$ is related to the mass $M_s$ contained in the Gaussian filter at mean density $\bar{\rho}$ via $M_s=(2\pi)^{3/2}\,\bar{\rho}R_s^{3}$; thus,  a smoothing scale $R_s=2.1\,h^{-1}\,{\rm Mpc}$ corresponds to $M_s=10^{13}\,h^{-1}\,{\rm M}_\odot$. 

A natural mass scale at any given redshift is given by the typical mass scale for collapse $M_\ast(z)$ defined as follows. A spherical top-hat perturbation collapses when its linear overdensity exceeds a value of $\delta_c=1.686$ with only a weak dependence on cosmological parameters \citep[e.g.][]{Eke1996}. The variance of linear density fluctuations at a given mass scale $M$ is related to the linear power spectrum $P(k,z)$ at redshift $z$ by
\begin{equation}
\sigma^2(M,z) = \frac{1}{2\pi^2}\int_0^{\infty}dk\,k^2\,P(k,z)\,\widetilde{W}^{2}_{\rm{TH}}(k,M),
\end{equation}
where $\widetilde{W}_{\rm{TH}}(k,M)$ is the Fourier transform of a spherical top-hat window function of comoving size $R=(3M\,/\,4\pi\bar{\rho})^{1/3}$, and $\bar{\rho}$ is the comoving mean mass density of the universe.  At a given redshift, the typical mass scale $M_\ast(z)$ to collapse from a $1\sigma$ fluctuation is hence given by the implicit solution of 
\begin{equation}
\sigma(M_\ast,z)=\delta_c.
\end{equation}
The  mass scale $M_\ast(z)$ at $z=0$ is $5.5\times10^{12}\,h^{-1}{\rm M}_\odot$.

When computing the environmental classification at redshifts $z>0$, there are two possible approaches that can be followed: {\it i)} To keep the smoothing scale $R_s$ ($M_s$) fixed to the $z=0$ value: the environment is thus defined over typical separations of a few Mpc in comoving space; or {\it ii)} To vary the smoothing scale. In particular, a natural choice for a redshift-dependent smoothing scale is to maintain the ratio $M_s/M_\ast$ fixed for the Gaussian filter. The respective mass scales $M_\ast(z)$ for the high-$z$ simulation snapshots investigated in this paper, i.e., $z=0.49$ and $z=1.05$, are $1.2\times10^{12}$ and $1.9\times10^{11}\,h^{-1}\,{\rm M}_\odot$. Fixing the ratio $M_s/M_\ast$ maintains the  smoothing on scales of order $\sim2M_\ast$ at all redshifts.

The resulting classifications for both $M_s={\rm const.}$ and $M_s/M_\ast={\rm const.}$ at redshifts $z=0$, $0.49$ and $1.05$ are shown in Figure \ref{fig:ScatterHaloes} using different colours for the cluster, sheet, filament and void  environments. We observe some differences between the two smoothing approaches. With a fixed smoothing length $R_s={\rm const.}$, shown in panels b) and d),  the regions classified as voids, sheets and filaments remain virtually unchanged since $z=1$, while the cluster environments grow substantially in size, from $z=0$ to higher redshifts, and extend well into the filaments at $z=1$. With the constant $M_s/M_\ast$ smoothing, shown in panels c) and e), a much larger number of individual haloes change environment with redshift:  at the resolution of our simulations, very few haloes are detected at $z=1$ in void regions, while many haloes are associated at the same redshift to relatively small cluster environments. In Tables \ref{matrix1} (for fixed $M_s$) and \ref{matrix2} (for fixed $M_s/M_\ast$) we show the fraction of haloes at $z=0$ that change  their environmental class from $z=1.05$ through  $z=0.49$ to $z=0$ by following the main progenitors of each halo with a minimum mass of $10^{11}\,h^{-1}\,{\rm M}_\odot$ in the $90\,h^{-1}\,{\rm Mpc}$ box. For a fixed smoothing mass scale, indeed less haloes change their environmental class as the density contrasts between the environments grow through gravitational collapse.

\begin{table}
\caption{\label{matrix1}Environmental transition matrix for the main progenitor branch of haloes with masses $M(z=0)>10^{11}\,h^{-1}\,{\rm M}_\odot$ between $z=0$ and $z=0.49$ (upper half) and $z=1.05$  (lower half). Matrix entries are given in per cent of the total number of haloes at $z=0$. Non-diagonal elements represent haloes that change classification. Environment is determined with $M_s={\rm const.}$ (i.e., $R_s=2.1\,h^{-1}\,{\rm Mpc}$ at all redshifts).}
\begin{center}
\begin{tabular}{@{}rcccc@{}}
\hline
& \multicolumn{4}{c}{z=0.49} \\
z=0 & {\it void} & {\it sheet} & {\it filament} &{\it  cluster} \\
\cline{2-5}
{\it void} & 1.0 & 0.094 & 0 & 0 \\
{\it sheet} & 0.28 & 27 & 1.1 & 0 \\
{\it filament} & 0 & 5.5 & 54 & 0.29 \\
{\it cluster} & 0 & 0.088 & 6.4 & 4.2 \\
\cline{2-5}
& \multicolumn{4}{c}{z=1.05} \\
z=0 & {\it void} & {\it sheet} & {\it filament} &{\it  cluster} \\
\cline{2-5}
{\it void} & 0.95 & 0.18 & 0 & 0 \\
{\it sheet} & 0.56 & 26 & 1.86 & 0 \\
{\it filament} & 0.028 & 10.3 & 49 & 0.46 \\
{\it cluster} & 0 & 0.53 & 7.5 & 2.7 \\
\hline
\end{tabular}
\end{center}
\end{table}

\begin{table}
\caption{\label{matrix2}As Table \ref{matrix1}, but with the environment now determined adopting  $M_s/M_\ast={\rm const}$.}
\begin{center}
\begin{tabular}{@{}rcccc@{}}
\hline
& \multicolumn{4}{c}{z=0.49} \\
z=0 & {\it void} & {\it sheet} & {\it filament} &{\it  cluster} \\
\cline{2-5}
{\it void} & 0.34 & 0.71 & 0.077 & 0 \\
{\it sheet} & 0.050 & 18 & 9.9 & 0.36 \\
{\it filament} & 0 & 6.2 & 51 & 2.9 \\
{\it cluster} & 0 & 0.31 & 8.0 & 2.4 \\
\cline{2-5}
& \multicolumn{4}{c}{z=1.05} \\
z=0 & {\it void} & {\it sheet} & {\it filament} &{\it  cluster} \\
\cline{2-5}
{\it void} & 0 & 0.22 & 0.79 & 0.12 \\
{\it sheet} & 0 & 5.0 & 19 & 4.1 \\
{\it filament} & 0 & 4.5 & 46 & 8.8 \\
{\it cluster} & 0 & 0.72 & 8.0 & 2.0 \\
\hline
\end{tabular}
\end{center}
\end{table}

\begin{figure}
  \begin{center}
    \includegraphics[width=0.35\textwidth]{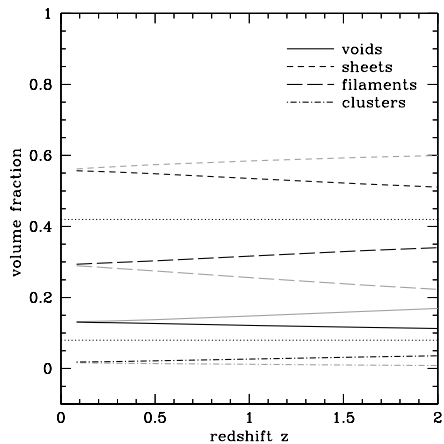}
  \end{center}
  \caption{ \label{fig:VolumeFractions} Volume fractions for the 90 $h^{-1}{\rm Mpc}$ box over redshift. Thick black lines represent the fractions obtained using a smoothing scale constant with redshift, and grey lines indicate the corresponding fractions obtained when keeping the ratio $M/M_\ast$ constant. Thin dotted black lines represent the values predicted for a Gaussian field (42 per cent for sheets and filaments, and 8 per cent for voids and clusters). }
\end{figure}

The fraction of  volume attributed to each of the four environments as a function of redshift is shown for both smoothing approaches  in Figure \ref{fig:VolumeFractions}. For a fixed smoothing mass $M_s={\rm const.}$, the density field asymptotically approaches Gaussianity with increasing redshift and thus the expected volume fractions for the four environments \citep[cf.][]{Doro70}. The behaviour is very different with the constant  $M_s/M_\ast$ smoothing. The volume occupied by the unstable structures (sheets and voids) increases with redshift, while the volume fractions of the stable structures (clusters and filaments) decrease compared to the values at $z=0$. 

\begin{figure}
  \begin{center}
    \includegraphics[width=0.35\textwidth]{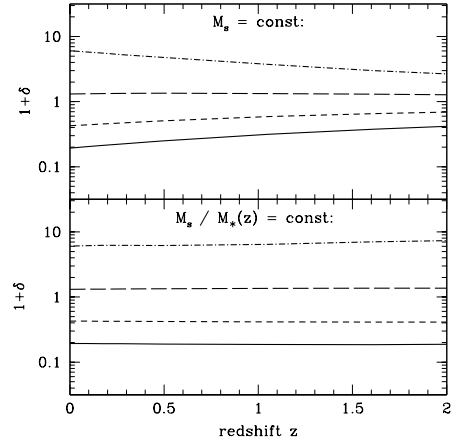}
  \end{center}
  \caption{ \label{fig:Densities} Median overdensity in cluster (dot-dashed), filament (long dashed), sheet (short dashed) and void (solid) environments for a constant smoothing scale $M_s$ (top) and for a constant ratio $M_s/M_\ast$ (bottom) with  redshift. The smoothing scale at redshift zero is $M_s=10^{13}\,h^{-1}\,{\rm M}_\odot$.}
\end{figure}

The redshift evolution of the median value of the smoothed overdensity field as measured at the centres of haloes in the four environments is shown in Figure \ref{fig:Densities}. For a constant smoothing scale, the median overdensities grow faster than expected in linear perturbation theory as $|\delta|\gtrsim 1$ in clusters and voids. With  the $M_s/M_\ast={\rm const.}$ smoothing, however, the median overdensities are found to be essentially constant at all redshifts. In both smoothing approaches, the median overdensity in filaments is constant. The median overdensities, smoothed on scales of $R_s=2.1\,h^{-1}\,{\rm Mpc}$ at redshift zero, are $\delta=-0.79$, $-0.55$, $0.28$ and $4.44$ in voids, sheets, filaments and clusters, respectively. 

We note that with both smoothing approaches,  and most relevantly when adopting a constant $M_s/M_\ast$ ratio for the smoothing, the haloes in the immediate surroundings of the clusters at $z=1$ are classified  as filaments/sheets at this redshift, but they make the transition to the cluster environment by $z=0$.  This allows us to rigorously identify, and thus study the properties of, the haloes in these intermediate-density ``transition regions'', before their ultimate migration into the deeper potential wells of rich clusters at $z=0$. We plan to investigate these haloes in future work.

\section{The Redshift Evolution of Halo Properties in Different Environments}
\label{sec:Haloes}

\subsection{Mass Functions}
\label{sec:MassFunction}

\begin{figure}
\begin{center}
  \includegraphics[width=0.35\textwidth]{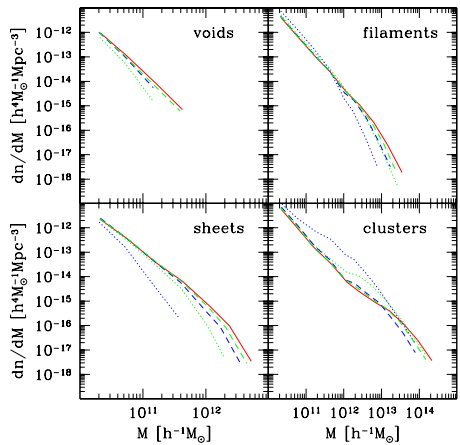}
\end{center}
\caption{ \label{fig:MassFunctions} Mass function of the haloes residing in voids, sheets, filaments and clusters at redshifts $z=0$ (red), $z=0.49$ (green) and $z=1.05$ (blue). Abundances in the whole box have been rescaled by the corresponding volume fractions. The dotted curves are obtained with $M_s/M_\ast={\rm const.}$, the dashed curves with $M_s={\rm const}$ at $z>0.49$ and $1.05$. The smoothing scale at redshift zero is $M_s=10^{13}\,h^{-1}\,{\rm M}_\odot$; here the two smoothings coincide and are represented by the solid line.}
\end{figure}

The choice of smoothing scale with redshift has an impact on the analysis of the redshift evolution of the halo properties in the different environments. Starting with the halo mass functions, shown in Figure \ref{fig:MassFunctions} for the cluster, sheet, filament and void environments at  $z=0$, $0.49$ and $1.05$, there is a substantial change in their shapes when using one or the other of the smoothing approaches. Adopting a constant  $M_s/M_\ast$  for the smoothing scale leads to a substantial increase in low mass haloes that are classified to be in clusters relative to the other environments. In voids, sheets and high mass filaments, the mass functions are higher when smoothing with $M_s={\rm const.}$ than when adopting a  constant  $M_s/M_\ast$ ratio; the trend reverses for haloes in low mass filaments and clusters, for which the mass functions are instead enhanced when using the $M_s/M_\ast={\rm const.}$ smoothing scale.  The inflexion point on scales of $\sim 10^{12}\,h^{-1}{\rm M}_{\odot}$  ($\sim 10^{12.5}\,h{-1}{\rm M}_{\odot}$) in the mass function of $z=1.05$ ($z=0.49)$ clusters for the $M_s/M_\ast={\rm const.}$  smoothing reflects the increasing abundance with redshift of isolated small clusters that we mention in Section \ref{sec:Classification}. At all redshifts of our study, the clusters have the highest mean number density of haloes, followed by filaments, sheets and voids, respectively. 
 
\subsection{Halo Formation Redshift}

\begin{figure*}
  \begin{center}
    \includegraphics[width=0.35\textwidth]{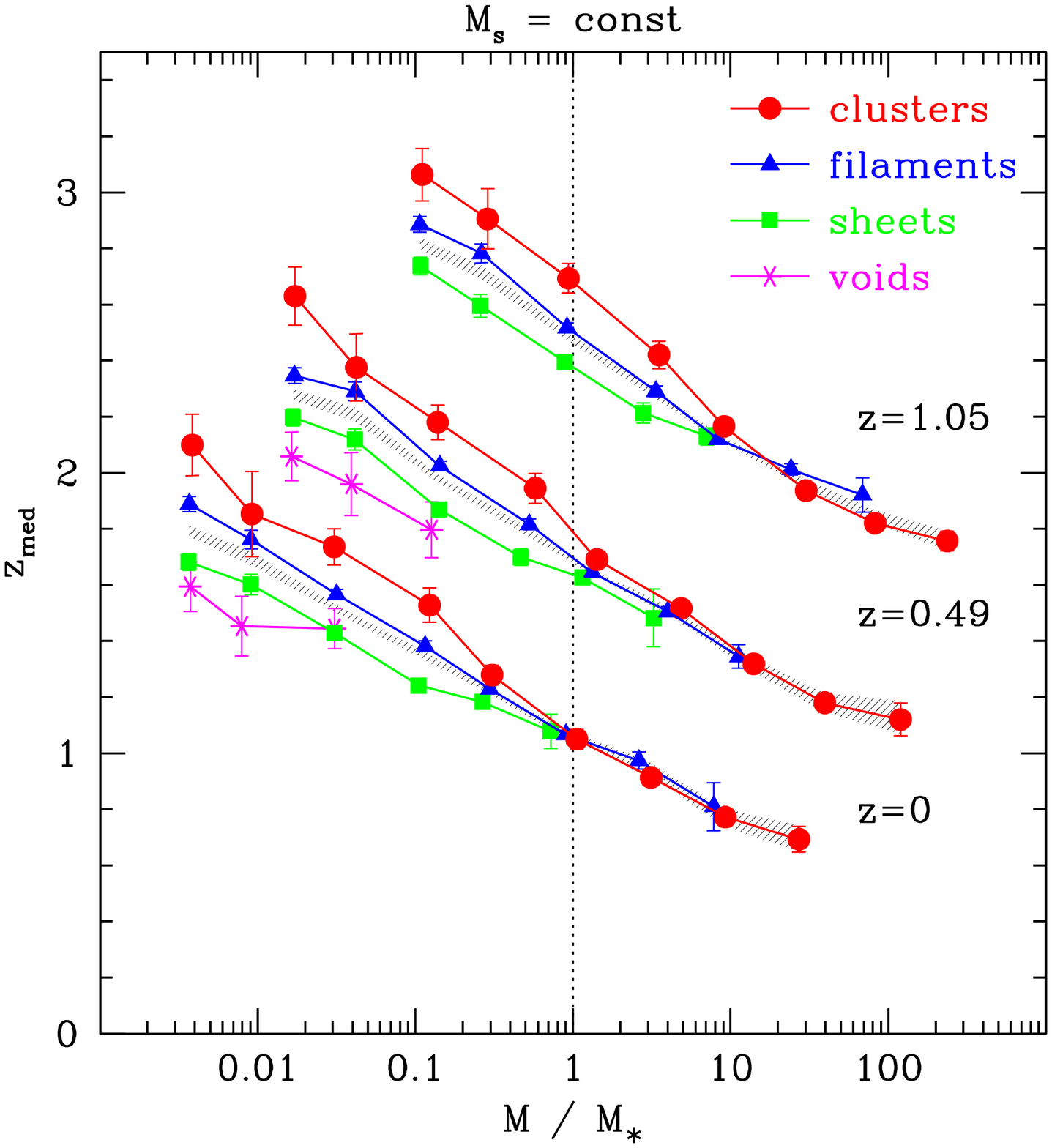}
    \hspace{1cm}
    \includegraphics[width=0.35\textwidth]{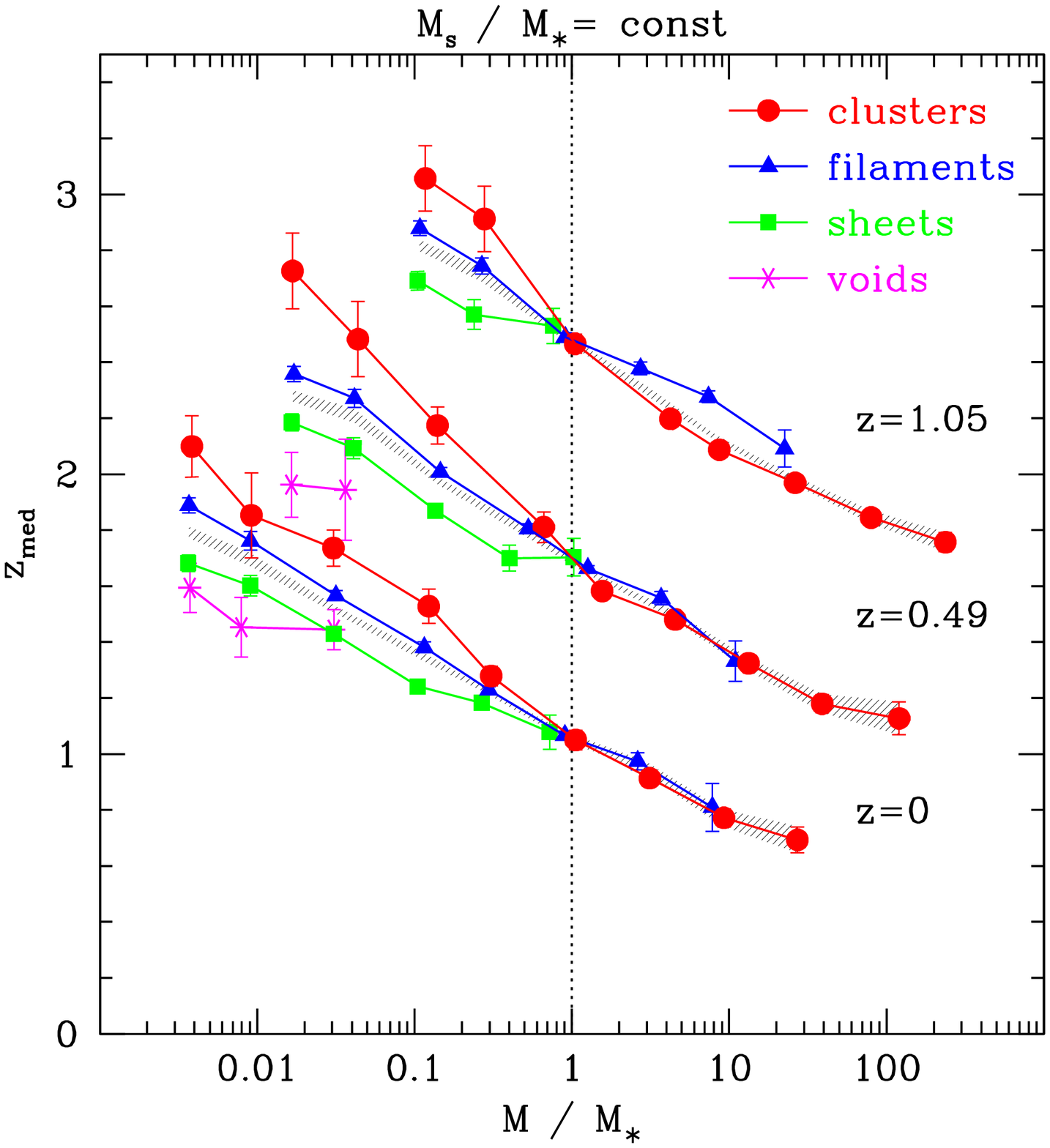}
  \end{center}
  \caption{ \label{fig:MassZForm} Median formation redshift in voids, sheets, filaments and clusters as a function of halo mass in units of $M_{\ast}(z)$ at redshifts $z=0$, $0.49$ and $1.05$. The left panel shows the results when smoothing with a constant $M_s$; the right panel shows the case of the smoothing with a constant  $M_s/M_\ast$ ratio. Errorbars indicate the uncertainty on the median. The shaded area indicates the 1$\sigma$ confidence area for the median of the whole sample not split by environment. }
  \end{figure*}
A closer inspection of the dependence of halo properties on the adopted smoothing scale shows that, with both approaches,  there is always a threshold mass-scale $M_{\rm t}$ below which the environmental influence on halo properties either switches sign or transitions from being negligible to becoming increasingly more substantial  with decreasing halo mass, down to the $\sim 10^{10}\,h^{-1}{\rm M}_{\odot}$ scales which can be probed with our simulations. Figure \ref{fig:MassZForm} shows, for both smoothing approaches,  the dependence on environment and mass (in units of $M_\ast$) of the median halo formation redshift $z_{\rm med}$. Errorbars are estimates of the error in the median, and are computed as:
\begin{equation}
\Delta x = \frac{x_{0.84}-x_{0.16}}{\sqrt{N_h}},
\end{equation}
where $N_h$ is the number of haloes used to sample the distribution of $x$, and $x_{0.84}$ and $x_{0.16}$ denote the 84th and 16th percentile of the distribution. These values would correspond to  $1\sigma$ errors if the underlying distribution were Gaussian. The $1\sigma$ confidence region of the median formation redshift determined from all environments is represented by the shaded regions in Figure \ref{fig:MassZForm}. As discussed in Paper I, this overall behaviour, not split by environment, is well fit by a logarithmic relation over five decades in mass at $z=0$. This relation, reflecting hierarchical assembly is also present at higher redshifts. We fit a model of the form
\begin{equation}
z_{\rm med} = c_1 - c_2 \times \log_{10} \frac{M}{M_{\ast}(z)},
\label{eq:ZFormFit}
\end{equation}
A least-squares fit to all haloes extracted from our three simulation boxes provides the following parameters at the three redshifts:
\begin{eqnarray*}
\left. \begin{array}{rrl}
c_1 & = & 1.07\pm 0.01, \\
c_2 & = & 0.30\pm 0.01;
\end{array} \right\} & & \vspace{-6pt}z=0,\\
\left. \begin{array}{rrl}
c_1 & = & 1.70\pm 0.01, \\
c_2 & = & 0.33\pm 0.01; 
\end{array} \right\} & & \vspace{-6pt}z=0.49,\\
\left. \begin{array}{rrl}
c_1 & = & 2.47\pm 0.01, \\
c_2 & = & 0.34\pm 0.01; 
\end{array} \right\} & & \vspace{-6pt}z=1.05.
\end{eqnarray*}
We note that, in all plots and thus at all redshifts and for both smoothing approaches, there is indeed a mass scale $M_{\rm t}$ at which the curves for the four environments meet, indicating the vanishing of significant environmental influence above this mass threshold. Specifically, below $M_{\rm t}$, haloes form earlier in clusters than in filaments, followed by sheets and then voids. This difference in formation time increases with decreasing mass below the threshold $M_{\rm t}$.

It is clear from Figure \ref{fig:MassZForm} that, in the case of a smoothing scale that remains constant  with redshift, the threshold $M_{\rm t}$ coincides with the mass-scale for gravitational collapse $M_\ast$ at $z=0$, but strongly departs (and increases relative to $M_\ast$) at higher redshifts.  Interestingly, however, in the case of the $M_s/M_\ast={\rm const.}$ smoothing scale, the threshold mass is easily identified to lie very close to $M_\ast$ at all redshifts.  This difference is simply due to the  different association of haloes to the {\it cluster} and {\it filament} environments in the two smoothing approaches that is also observed in the mass functions (cf. Figure \ref{fig:MassFunctions}).
Furthermore, we note that the $z=0.49$ and $z=1.05$ relations are very similar to the one at $z=0$, for which we had already provided the analytic fits with environment-dependent slopes for masses $M<M_\ast$ in Paper I. The environmental dependence of the halo formation redshifts below the $M_\ast$ mass scale that we have found agrees with the analysis of \cite{Gao06}, who find that haloes with higher formation redshifts are more strongly clustered.

Interestingly, at the highest redshift of our study ($z=1.05$),  and for the $M_s/M_\ast={\rm const}$ smoothing, our simulations  show that haloes with masses $M\gtrsim M_\ast$ in filaments tend to have earlier formation times  than haloes of similar masses in the cluster environment, i.e., an opposite trend than the one observed at all redshifts below $M=M_\ast$. 

The fact that, in the case of a fixed $M_s/M_\ast$ smoothing ratio, the environmental dependence is explicitly seen to appear around $M/M_\ast=1$, hints at a physical relevance of this scale in establishing the onset of the environmental dependence of halo properties at all redshifts. This motivates us to identify the constant $M_s/M_\ast$ ratio as the more fundamental smoothing scale in our analysis, and thus to use this smoothing scale in the remainder of our study of the redshift evolution of halo spins, shapes and alignments as a function of environment.

\subsection{Halo Spin}
\begin{figure}
  \begin{center}
    \includegraphics[width=0.35\textwidth]{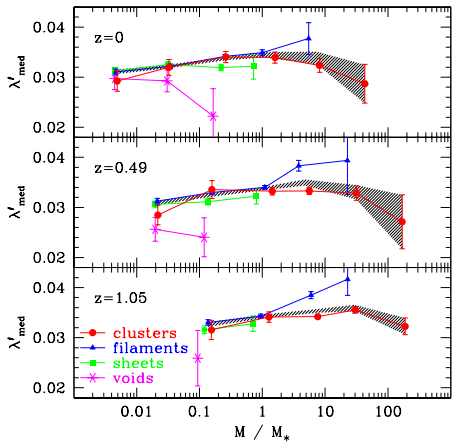}
  \end{center}
  \caption{ \label{fig:MassSpin} Median spin parameter in filaments and clusters as a function of halo mass in units of $M_{\ast}$ at redshifts $z=0$, $z=0.49$ and $z=1.05$. Errorbars indicate the uncertainty on the median. The shaded grey area indicates the 1$\sigma$ confidence area of the median of the whole sample not split by environment.}
\end{figure}

Figure \ref{fig:MassSpin} shows the median spin parameter $\lambda^\prime_{\rm med}$ as a function of normalised mass $M/M_\ast$ for haloes at $z=0$, $0.49$ and $1.05$ in the cluster, sheet, filament and void environments. 
  We also plot the results integrated over all environments (shaded region in Figure \ref{fig:MassSpin}), for comparison with previous studies. In agreement with these  \cite[e.g.][]{Vitvitska2002}, we do not observe any significant evolution of the global spin parameter  with redshift. Our results also confirm a weak mass dependence of the median spin parameter as found at $z=0$ by \cite{Bett06}, and furthermore extend this result to significantly lower masses. Integrated over all environments, the spin-mass relationship is weakly increasing up to $\approx 10M_\ast(z)$, while the highest masses at each redshift have again a slightly smaller median spin parameter.

We find some dependence of the halo spin on environment at all redshifts of our analysis. More specifically, high-mass ($M>M_\ast$) haloes in filaments have a higher median spin at all redshifts than comparable-mass haloes in the cluster environments; in the voids, haloes with masses substantially below $M_\ast(z)$  spin systematically more slowly (i.e. median $\lambda^{\prime}\lesssim 0.03$) than haloes of similar masses in any other environment. We have tested that this trend of void haloes having lower median spin parameters persists and actually increases when a larger scale is adopted for the smoothing, to optimize the identification of the void regions (cf. Paper I). At the lowest masses we do not observe any significant difference between haloes in clusters, filaments or sheets.

\cite{Gao06} report that, in their simulations, the most rapidly spinning haloes are more clustered than the slowest spinning haloes, which is in agreement with our earlier findings at $z=0$ (Paper I). Our results, after removing unrelaxed haloes as described in Section \ref{sec:HaloCatalogues}, do not support a very strong correlation between environment and spin at low masses. Still, it is slightly more likely to find the most rapidly spinning objects in environments of higher median density.

\subsection{Halo Shape}

\begin{figure*}
  \begin{center}
    \includegraphics[width=0.35\textwidth]{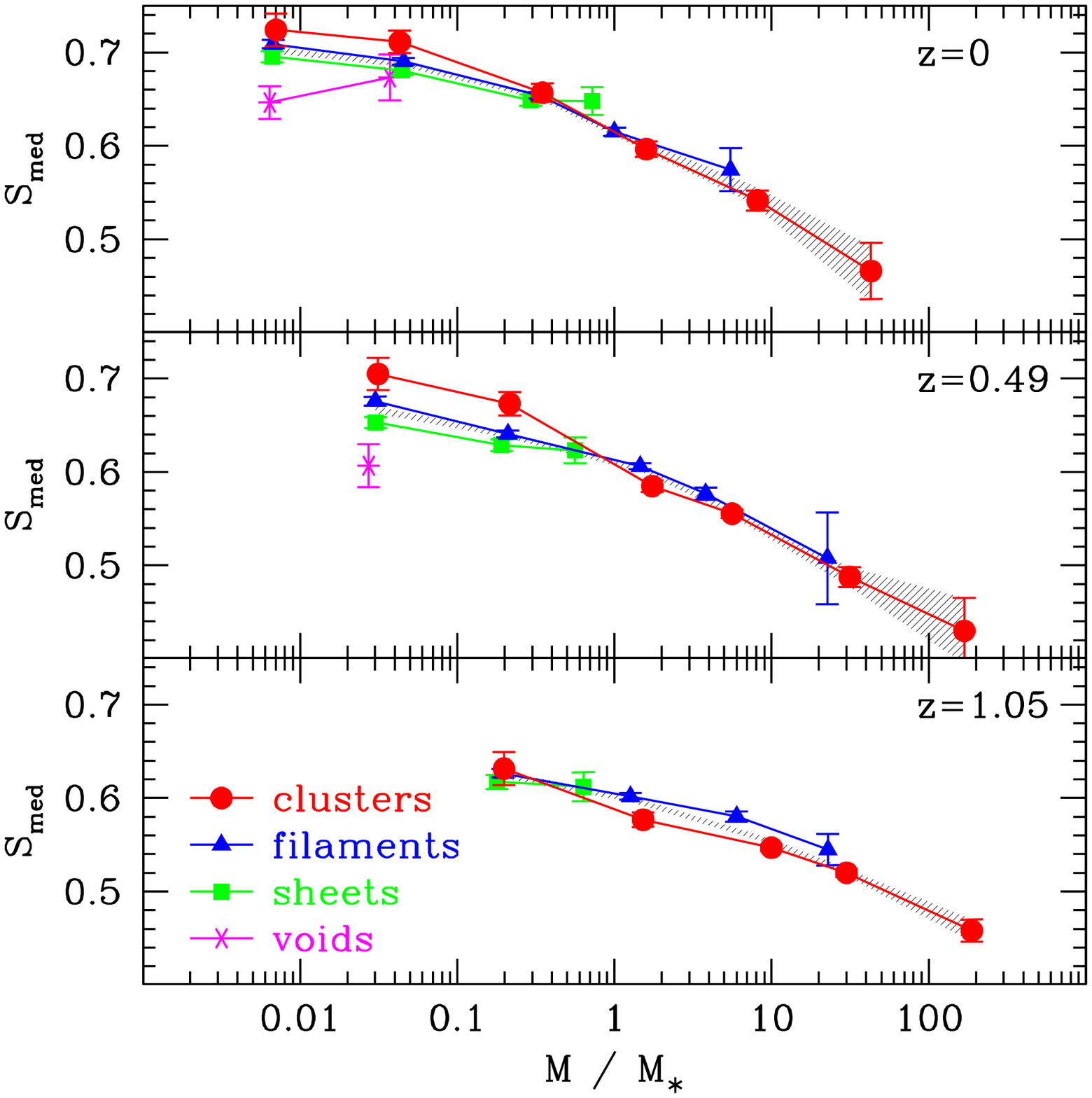}
    \hspace{1cm}
    \includegraphics[width=0.35\textwidth]{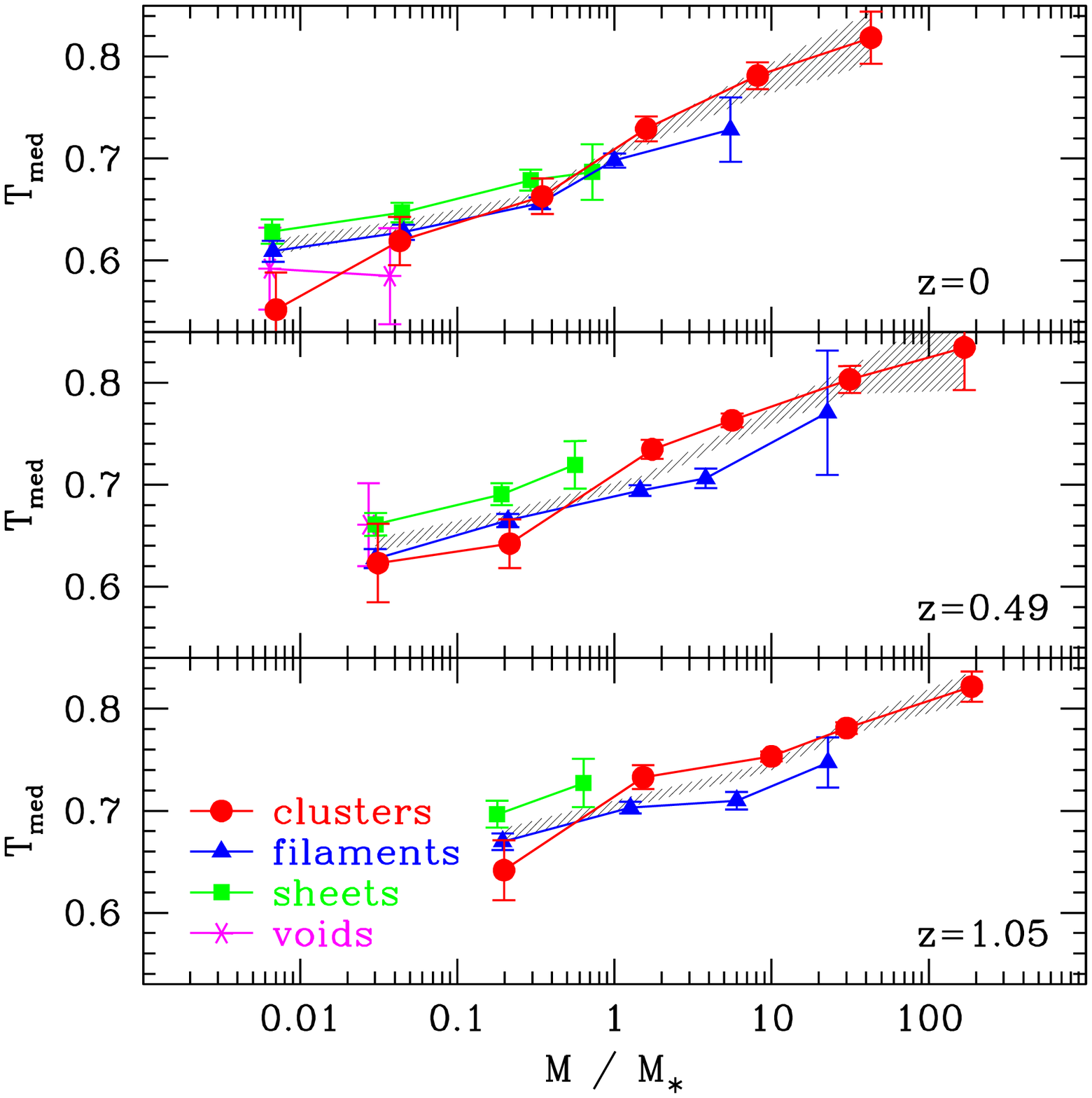}
\end{center}
\caption{ \label{fig:HaloShapes} Median halo sphericity (left) and triaxiality (right) as a function of halo mass in units of $M_\ast$ for haloes in the four environments at redshift $z=0$, $0.49$ and $1.05$. Environment is defined adopting a constant  $M_s/M_\ast$ ratio. Errorbars represent the error in the median. The shaded grey area indicates the 1$\sigma$ confidence area of the median of the whole sample not split by environment.}
\end{figure*}

In Figure \ref{fig:HaloShapes}, we show the median sphericity and triaxiality of haloes in the four environments as a function of their mass for redshifts $z=0$, $0.49$ and $1.05$. Independent of redshift and environment,  haloes  tend to be more spherical with decreasing mass. Over a large range of masses, haloes at $z=1.05$ are however less spherical than haloes of similar mass at $z=0$; \cite{AvilaReese2005} and \cite{Allgood06} find a similar result in their analyses. The mass-shape relations, expressed as scale-free functions of $M/M_\ast$, show no significant evolution with redshift up to $z\sim1$; the fact that the entire redshift evolution of the shapes of haloes is driven by the evolution of the mass scale for gravitational collapse,  $M_\ast$, is also supported by the independent studies quoted above.

Similarly to the  $z=0$ case \citep[][Paper I]{Bett06},  the mass-sphericity and the mass-triaxiality relation of the global (i.e., not split for environment) sample follow a broken logarithmic relation also at high redshifts. The change in slope in these relationships occurs around $M\approx M_\ast$ . We detect however  a  relatively small but systematic  difference in the sphericity and triaxiality of  $M<M_\ast$ halos in low- and high-density environments, i.e., the slope in the $M\lesssim M_\ast$ regime weakly depends on the environment.
Specifically, the median sphericity of $M<M_\ast(z)$ haloes decreases systematically from  the cluster environments, to the filaments, sheets and voids. At all redshifts, a weak trend is observed for haloes with masses below $M_\ast$ to be more oblate in clusters than in filaments, and more prolate in sheets than in filaments; for haloes above $M_\ast$, there is a stronger evidence for  haloes in filaments to be systematically more oblate than in clusters. These environmental differences at low masses are observed to be already in place at $z=0.49$; the resolution of our simulations is not adequate to properly investigate these effects at $z=1.05$ (minimum halo mass $\approx0.1M_\ast$).

\subsection{Halo Alignments}
\subsubsection{Halo-LSS alignment}

\begin{figure}
  \begin{center}
    \includegraphics[width=0.4\textwidth]{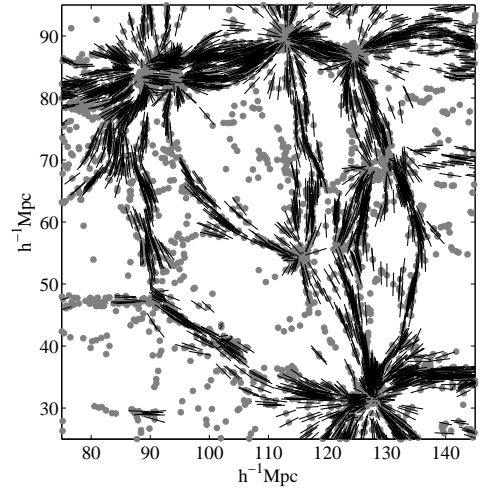}
  \end{center}
  \caption{ \label{fig:QuiverFilament} Unit eigenvectors indicating the direction of the filaments are shown in black for filament haloes in a slice of 8 $h^{-1}\,{\rm Mpc}$ in the 180 $h^{-1}\,{\rm Mpc}$ box at z=0. The grey symbols indicate halo positions regardless of their environment. The directional information of these vectors is used to determine the  alignment of halo spins with the large-scale structure.}
\end{figure}

\begin{figure}
  \begin{center}
    \includegraphics[width=0.35\textwidth]{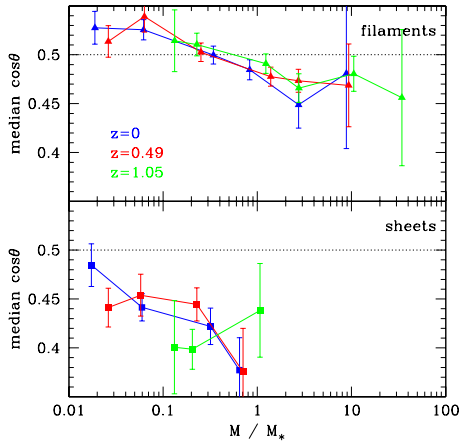}
  \end{center}
  \caption{ \label{fig:StructAlign} Median alignment angles between the halo angular momentum vectors and the eigenvectors pointing in the direction of filaments and normal to the sheets, respectively. Different redshifts are indicated with the three colours. Errorbars indicate the error in the median. The dotted line indicates the expectation value for a random signal.} 
\end{figure}

Extending the analysis  of Paper I to redshifts above zero, we use the directional information derived from the eigenstructure of the tidal field tensor to estimate the alignment of halo spin with the dynamical properties of the surrounding environment.  Filaments and sheets have a preferred direction given by the eigenvector corresponding the single positive or negative eigenvalue. The eigenvectors indicating the direction of the filament as determined from the tidal tensor are shown in Figure \ref{fig:QuiverFilament}. Given these unit eigenvectors $\hat{\mathbf{v}}$, we compute the alignment angle $\cos \theta = {\hat{\mathbf{J}} \cdot \hat{\mathbf{v}}}$.
Figure \ref{fig:StructAlign} shows the median alignment as a function of mass at redshifts $z=0$, $0.49$ and $1.05$. At all redshifts, there is a strong tendency for sheet haloes to have a spin vector preferentially parallel to the sheet, i.e. orthogonal to the normal vector. At redshifts up to $0.49$, where the errorbars of our measurements allow us to investigate  trends with halo mass,   this alignment increases with increasing mass. For filament haloes, there is a clear trend with halo mass: {\it (i)} haloes with masses smaller than about $0.1M_\ast$ have spins more likely aligned with the filament in which they reside; {\it (ii)}  haloes in the range $M\approx 0.1M_\ast$ to $1M_\ast$ appear to be randomly aligned with respect to the large-scale structure; and {\it (iii)}  For $M\gtrsim M_\ast$, the trend appears to reverse, and more massive haloes have a weak tendency to spin orthogonally to the direction of the filament at lower redshifts\footnote{The tendency for haloes above $M_\ast$ to spin orthogonal to the host filament, shown  in Figure \ref{fig:StructAlign} for the $M_s/M_\ast={\rm const.}$ smoothing case, is  enhanced when the $M_s={\rm const.}$ smoothing is adopted.  The smoothing scale not only determines the environmental split of the halo population, it also affects the scale on which the eigenvectors of the tidal field are computed. When the smoothing is performed with $M_s/M_\ast={\rm const.}$, the filament direction is obtained on increasingly smaller comoving scales at higher redshifts. This partially erases the stronger correlation that is observed for the most massive haloes when the smoothing is kept at constant comoving scale for all redshifts.}.

\begin{figure}
  \begin{center}
    \includegraphics[width=0.35\textwidth]{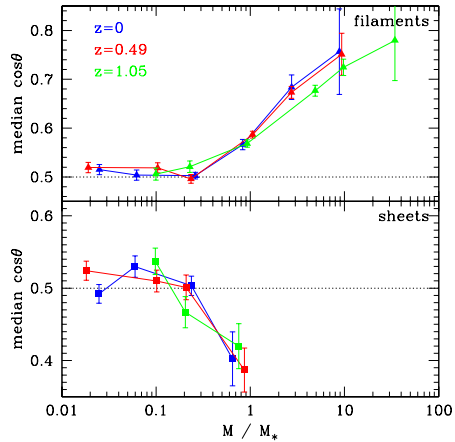}
  \end{center}
  \caption{ \label{fig:StructAlignShape} Median alignment angles between the halo  major axis vectors and the eigenvectors pointing in the direction of  filaments and normal to the sheets, respectively. Different redshifts are indicated with the three colours. Errorbars indicate the error in the median. The dotted line indicates the expectation value for a random signal. Data is shown for the ratio of the smoothing scale $M_s/M_\ast$ fixed.}
\end{figure}

To further explore possible connections  between  the alignment of the large-scale structure and the intrinsic alignment of haloes in the different environments,   we search for a correlation signal between the LSS and the axis vectors of the moment of inertia ellipsoid of the haloes. In particular, we use the major axis vector $\mathbf{l}_1$ to define the alignment angle $\cos \theta = {\hat{\mathbf{l}}_1 \cdot \hat{\mathbf{v}}}$, where $\mathbf{v}$ is again the eigenvector normal to a sheet or parallel to a filament. The resulting median correlation is shown in Figure \ref{fig:StructAlignShape}. We find no alignment for halo masses $M<0.1M_\ast$;  however,  in  both the filaments and the sheets, the halo major axis appears to be strongly aligned with the LSS for masses above about a tenth of $M_\ast$. The strength of the alignment grows  with increasing mass. This is possibly to be expected, especially for the most massive haloes, since  their shape might influence the potential from which the eigenvectors are derived. Adopting a fixed smoothing scale $M_s$ results merely in a shift of the relations shown in Figure \ref{fig:StructAlignShape}.

Results similar to ours concerning the alignments of shapes and spins with the LSS, and the transition of alignment orientation at $M_\ast$ in the filaments,  are reported by \cite{Aragon06} for $z=0$ haloes using a definition of environment that is based on density rather than, as in our case, on the gravitational potential, as well as for haloes in the vicinity of clusters by \cite{Basilakos2006} using the moment of inertia ellipsoid of superclusters and by \cite{Ragone2007} defining environment by the distance to the nearest cluster. It is clear from our present analysis that such alignments are in place at redshifts of order one, and are maintained virtually unchanged over the last eight or more billion years of evolution of structure in the universe.

\subsubsection{Halo-Halo alignments}

\begin{figure*}
  \begin{center}
    \includegraphics[width=0.3\textwidth]{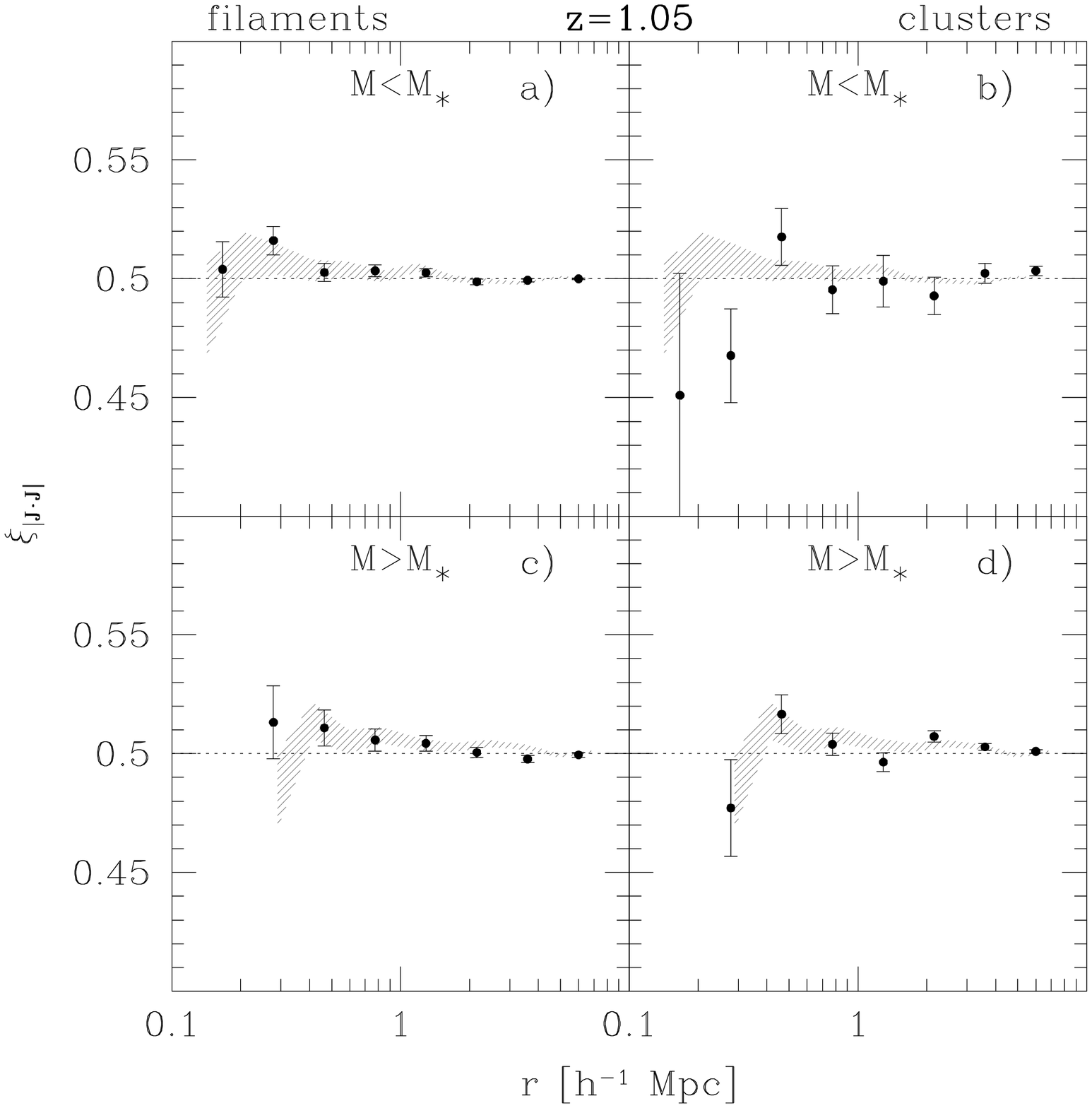}
    \includegraphics[width=0.3\textwidth]{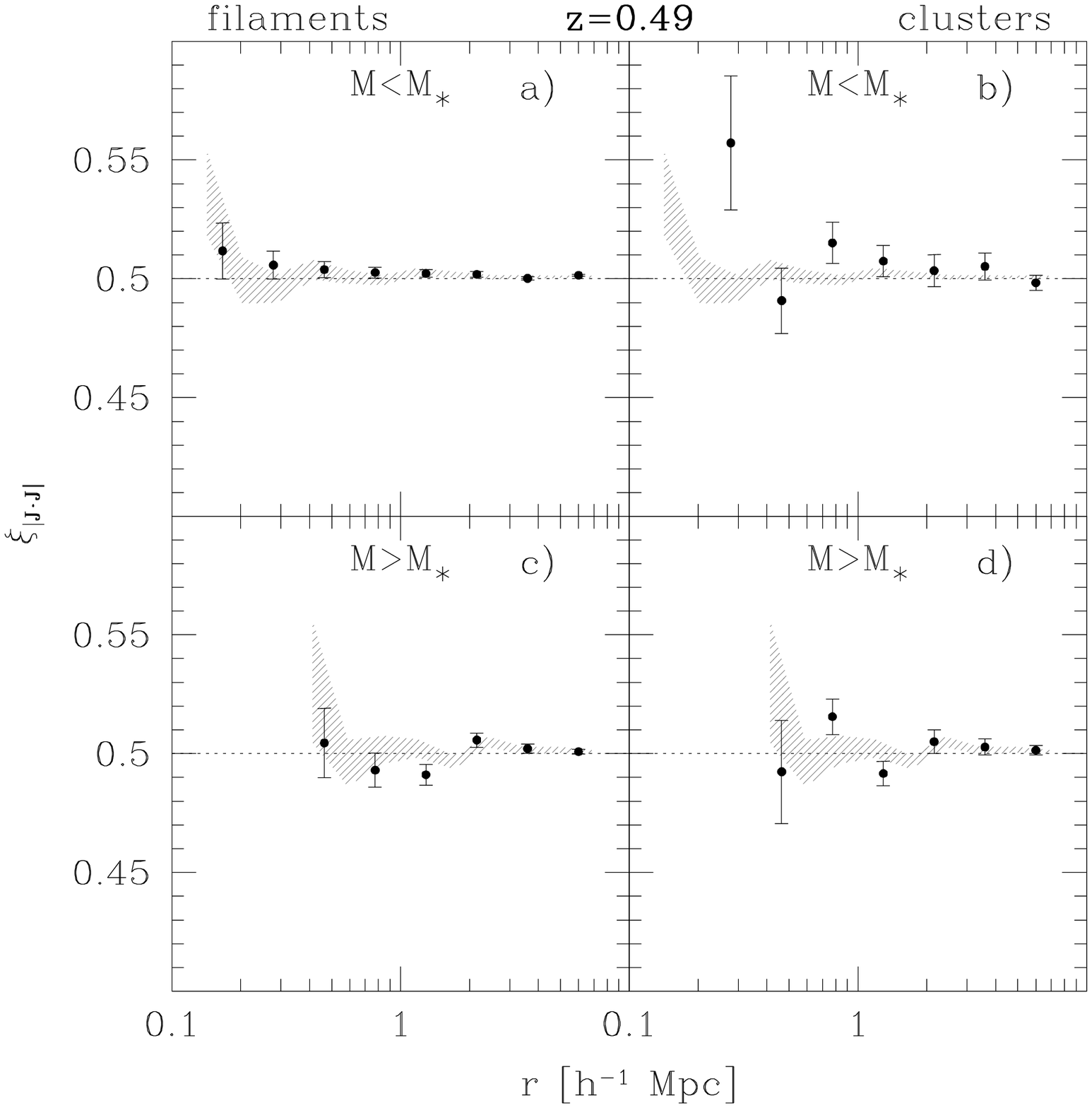}
    \includegraphics[width=0.3\textwidth]{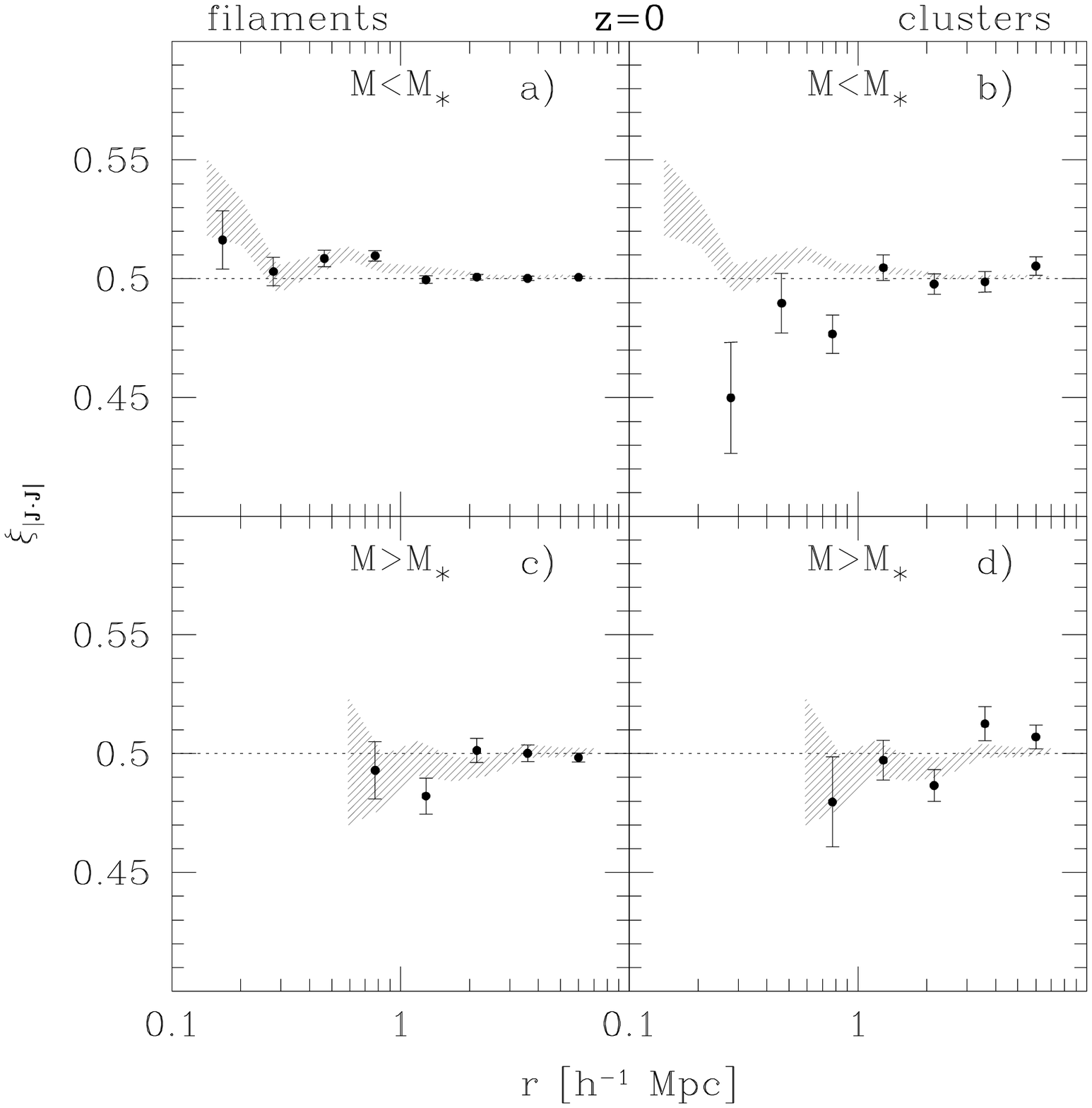}
  \end{center}
  \caption{ \label{fig:AlignJJ_var} The mean alignment of intrinsic spin angular momentum between haloes in filaments and clusters at redshifts $z=1.05$ (left), $z=0.49$ (centre) and $z=0$ (right). Data for filaments are shown in panels a) and c), clusters in panels b) and d). The upper panels a) and b) show the results for halo masses $M<M_\ast$, and the lower panels c) and d) for halo masses $M>M_\ast$. The dotted line indicates the expectation value for a random uncorrelated signal. The shaded region indicates the $1\sigma$ confidence interval on the mean for the whole sample, split by mass but not split by environment.}
\end{figure*}

\begin{figure*}
  \begin{center}
    \includegraphics[width=0.3\textwidth]{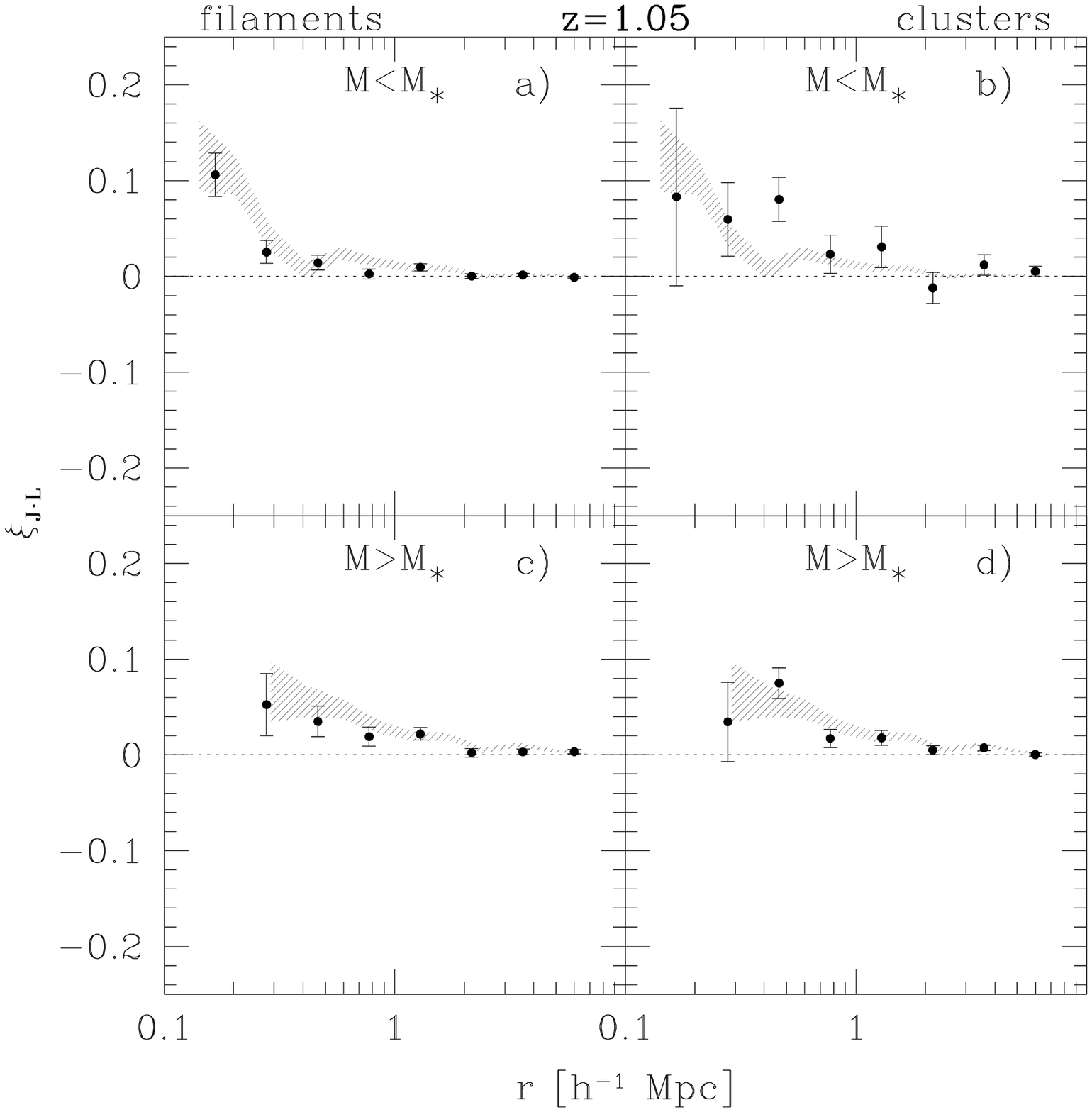}
    \includegraphics[width=0.3\textwidth]{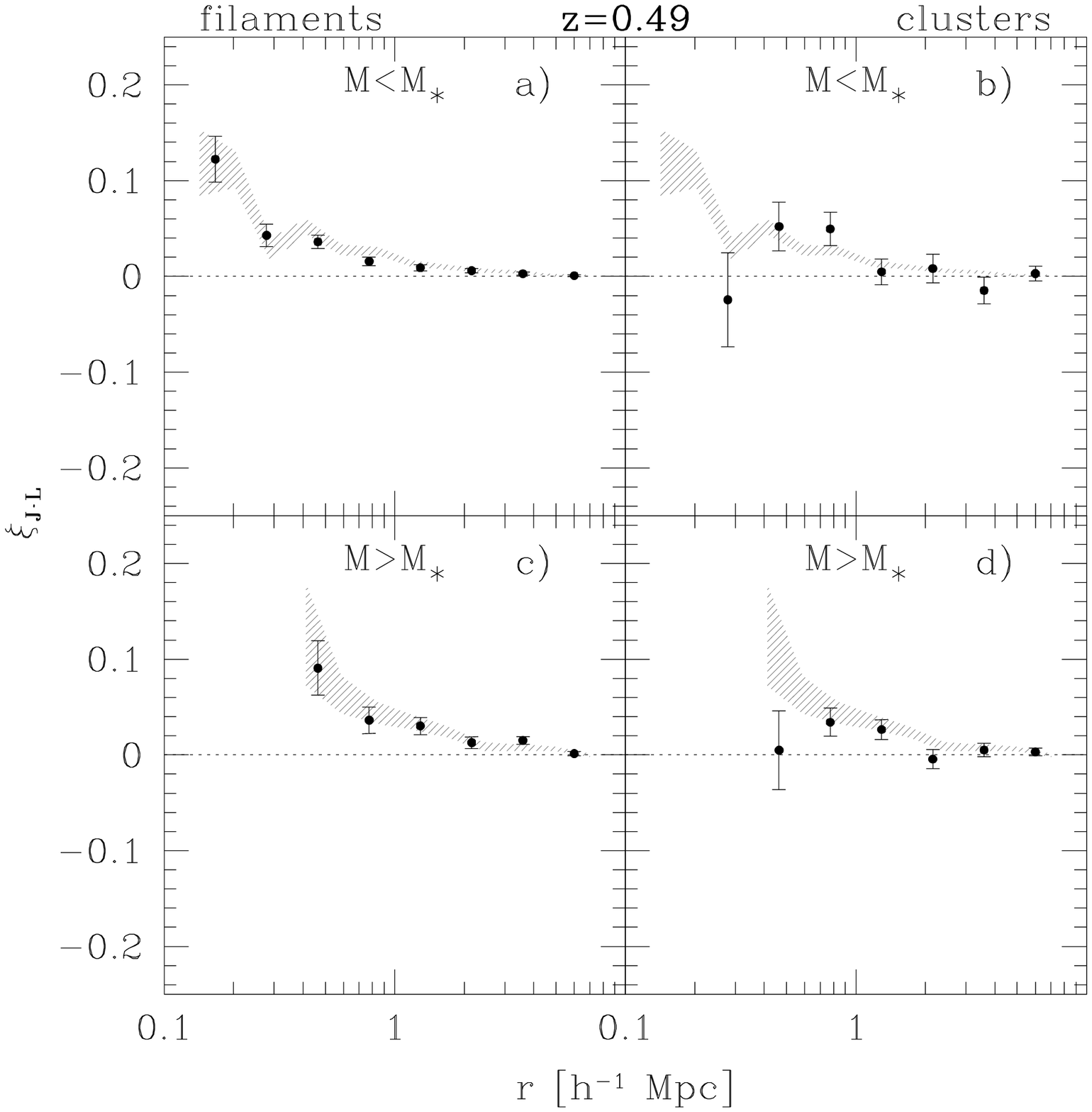}
    \includegraphics[width=0.3\textwidth]{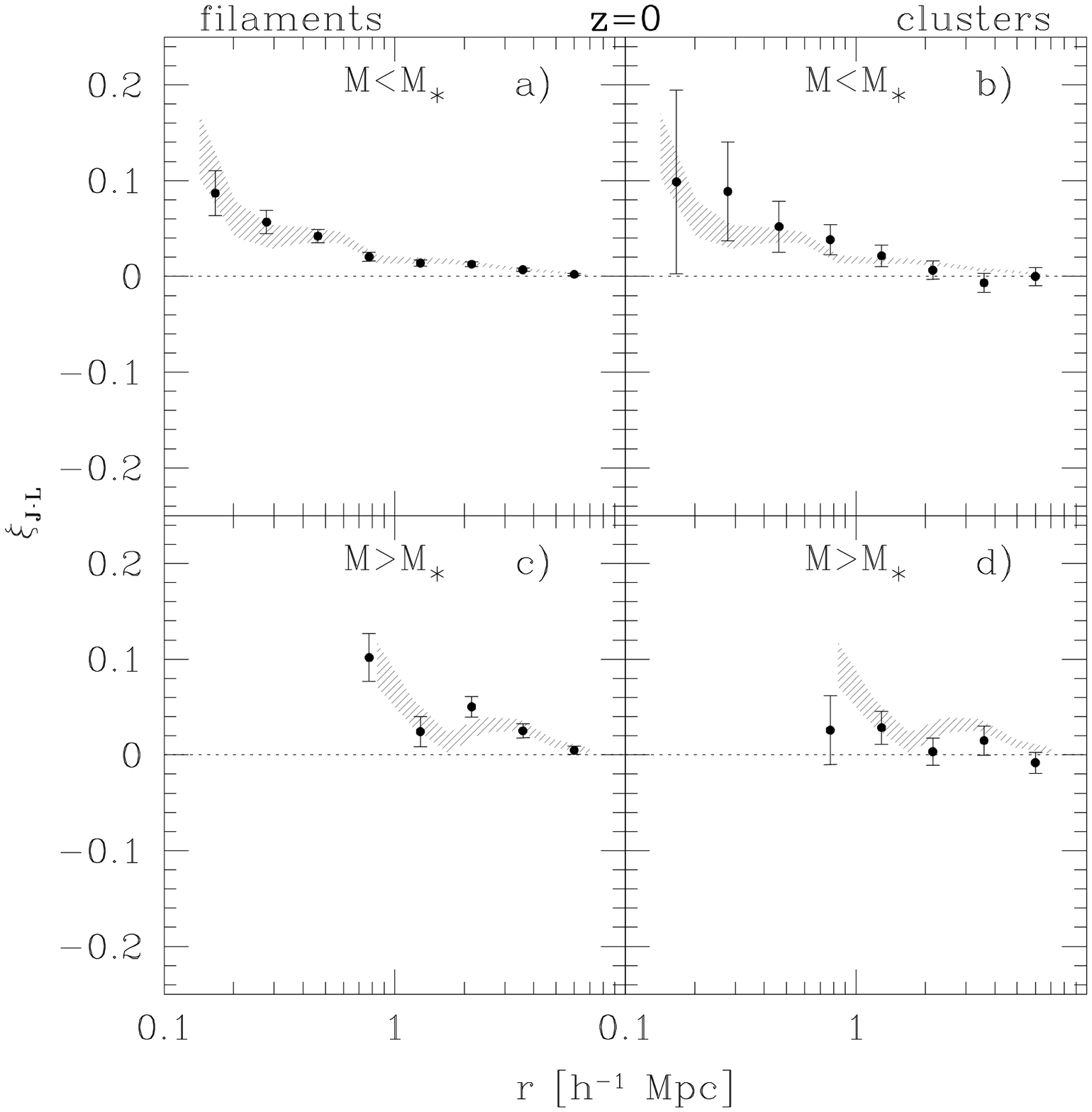}
  \end{center}
  \caption{ \label{fig:AlignJL_var} The mean alignment of intrinsic spin and relative orbital angular momentum between haloes in filaments and clusters at redshifts $z=1.05$ (left), $z=0.49$ (centre) and $z=0$ (right).  Data for filaments are shown in panels a) and c), clusters in panels b) and
d). The upper panels a) and b) show the results for halo masses $M<M_\ast$, and the lower panels c) and d) for halo masses $M>M_\ast$. The dotted line indicates the expectation value for a random uncorrelated signal. The shaded region indicates the $1\sigma$ confidence interval on the mean for the whole sample, split by mass but not split by environment.}
\end{figure*}

We finally compute the spin-spin and spin-orbit correlation functions using
the definitions  of  \cite{Porciani02} and \cite{Bailin2005}.  While we show the results for the $M_s/M_\ast={\rm const.}$ smoothing case, we stress that qualitatively the results remain unchanged when the constant smoothing is adopted.

For the spin-spin correlation we have:
\begin{equation}
\xi_{\mathbf{J}\cdot\mathbf{J}}(r) = \langle \,\arrowvert \hat{\mathbf{J}}({\bf x})\cdot\hat{\mathbf{J}}({\bf x}+{\bf r}) \arrowvert\,\rangle,
\end{equation}
where $\mathbf{J}$ is the intrinsic angular momentum of each halo, and the average is taken over all pairs of haloes which are separated by a distance $r$ and reside in the same environment class.
Similarly, the spin-orbit correlation is defined as:
\begin{equation}
\xi_{\mathbf{J}\cdot\mathbf{L}}(r) = \langle \hat{\mathbf{J}}({\bf x})\cdot\hat{\mathbf{L}}({\bf x}+{\bf r})\rangle,
\end{equation}
where $\mathbf{L}$ is the relative orbital angular momentum between two haloes separated by a distance $r$.

Figure \ref{fig:AlignJJ_var} shows the spin-spin alignment for haloes in clusters and filaments at the three redshifts of our study; upper and lower panels show respectively the results for  haloes with masses below  and above $M_\ast$. The shaded region shows the $1\sigma$-confidence area for the total sample, split by mass but not split by environment. The correlations within either of the environmental classes is never stronger than those for the total sample and all of them are consistent with no signal within $2\sigma$. Furthermore, we find no evidence for any significant redshift evolution of these correlations.

The spin-orbit correlation function is shown in Figure \ref{fig:AlignJL_var}. The strong correlation that we found at $z=0$ in  Paper I, extending out to several Mpc, is present also out to redshift $z=1$ with no significant changes.

\section{Summary and Conclusions}
\label{sec:Conclusion}

We have used three N-body simulations, tailored to cover a range of almost five decades in mass with high resolution haloes ($>300$ particles), to investigate the dependence of halo shape, spin, formation redshift and alignment as a function of mass, environment and redshift.  Using the tidal stability criterion of Paper I we have classified haloes to reside in four different environments: {\it clusters}, {\it filaments}, {\it sheets} and {\it voids}. The attribution of haloes to these environments depends on one free parameter, $R_s$, the length scale used to smooth the underlying mass distribution. Relating this length scale to the mass contained in the Gaussian filter, $M_s$, in  Paper I we optimised by visual inspection the redshift zero value of $M_s=10^{13}\,h^{-1}{\rm M}_\odot\approx 2M_\ast(0)$, with $M_\ast(z)$ the typical mass scale collapsing gravitationally at redshift $z$. 
At the higher  redshifts that we study in this paper, we discuss two possible choices for the smoothing mass scale: {\it i)} a smoothing scale constant with redshift; and {\it ii)} a smoothing scale that varies such that $M_s/M_\ast$ remains constant with redshift. 
The first approach leads to the median overdensity in each environment increasing just as expected from non-linear enhancement of density fluctuations; the second approach maintains the median density in each environment constant with redshift. 

In our analysis of the redshift evolution of the halo properties we find that, when adopting a constant ratio $M_s/M_\ast$ for the smoothing, the environmental influence is roughly invariant with redshift so that the mass scale at which the environmental influence sets in is roughly given by by the mass scale $M_\ast$. Unveiling the importance of this physical mass scale in the onset of an environmental dependence of the halo properties is a first step towards understanding the origin of the environmental role in the evolution of dark matter haloes.

Adopting the physically-motivated  $M_s/M_\ast={\rm const.}$ smoothing at all redshifts, we have investigated the dependence of the properties of isolated dark matter haloes of masses below and above the $M_\ast$ threshold on their environment.  In general, we find that all halo properties show some dependence on environment for halo masses $M\lesssim M_\ast$. The strength of the correlations, however, does not change much with redshift. There is virtually no redshift evolution of the halo properties when the correlations with halo mass are expressed in terms of the normalised mass $M/M_\ast$, indicating that the strongest evolution with redshift is related to the evolution of the mass scale for collapse, $M_\ast$. In detail, our main results are summarised as follows:
\begin{itemize}
\item{} There is a strong environmental dependence of halo formation times with environment for haloes with masses $M\lesssim M_\ast$. At any given mass in this mass regime, haloes in clusters tend to be older than haloes in the other environments, and haloes in voids form much more recently than in any other environment. The global halo population (with no splitting for environment) is well described by a logarithmic relation between mass and formation redshift with a roughly constant slope with redshift.
\item{} The median spin parameter $\lambda^\prime_{\rm med}$ of the total halo sample, not split by environment, increases weakly with mass up to around $10M_\ast$ at all redshifts. There is no significant residual dependence on redshift besides the mass rescaling with  $M_\ast$. There is an additional tendency for high mass haloes in filaments to spin more rapidly than haloes of the same mass in clusters. Haloes in voids have the lowest median spin parameters.
\item{} Haloes of a given ratio $M/M_\ast$ have very similar median shape parameters independent of redshift in the mass and redshift ranges that we have investigated. Independent of environment, haloes are increasingly more spherical and less triaxial the lower their mass. Haloes with masses $M\lesssim M_\ast$  are slightly more spherical and more oblate in clusters than in filaments, and there is a hint that the situation reverses for $M\gtrsim M_\ast$, i.e., that haloes in filaments are more oblate than cluster halos at high masses. Low-mass haloes in voids have systematically the lowest median sphericity of similar mass halos in denser environments.
\item{} In the $M\lesssim M_\ast$ mass regime, haloes in sheets tend to have spin vectors in the plane of the sheets, and haloes in filaments tend to have spin vectors pointing along the filaments;  above the $M_\ast$ mass scale, there is evidence that haloes in filaments reverse the previous trend and tend to have spins orthogonal to the filaments. Furthermore, haloes  with masses $M>0.1M_\ast$ tend to have their major axis parallel to their host sheets or filaments, with the strength of the alignment increasing with increasing mass.   This may reflect the fact  that, for massive haloes, the gravitational potential field is substantially influenced by their shape and thus leads to an aligned tidal field. The alignment of halo spins and major axes with the large-scale structure that we have unveiled up to redshifts of order $z=1$ should be taken into account in studies of weak lensing maps of cosmic shear \citep[eg.][]{Catelan2001}, especially in sheets and thus in regions surrounding voids.
\item{} There is no evidence for a significant spin-spin correlation between neighbouring haloes. There is in contrast a substantial halo spin-orbit alignment, whose strength appears to be independent of mass, environment and redshift up to $z\sim1$: haloes in close pairs tend to spin preferentially parallel to the orbital angular momentum of the pair.
\end{itemize}
An important conclusion that we draw from our study is that the environmental influence on halo properties shows an intriguing  dependence on the halo mass, and appears to be essentially modulated by the typical mass scale of gravitational collapse $M_\ast$ at each redshift. Our data suggests that the $M=M_\ast(z)$  mass scale  might indeed play the role of a {\it bifurcation point} below which many of the median properties of dark matter haloes  either begin to feel the influence of their large-scale environment, or show an opposite response to their large-scale environment relative to the more massive haloes. The existence of such a thresholding mass scale in the environment-halo relationship is yet to be understood. 
\section*{Acknowledgements}
OH acknowledges support from the Swiss National Science Foundation.
All simulations were performed on the Gonzales cluster at ETH Zurich, Switzerland.

\label{lastpage}

\end{document}